\documentclass[aps,pra,twocolumn,10pt]{revtex4-1}
\usepackage{graphicx}
\usepackage{amsmath}
\usepackage{amssymb}
\usepackage{mathrsfs}
\usepackage{amsfonts}
\usepackage{dsfont}
\usepackage{dcolumn}
\usepackage{bm}
\usepackage{booktabs}
\usepackage{tabularx}
\usepackage{textcomp}
\usepackage{color}
\usepackage{xcolor}

\everymath{\displaystyle}

\newcommand{\vE}{\mathbf{E}}

\newcommand{\di}{\mathrm{d}}
\newcommand{\hbe}{\hat{\bm{e}}}

\newcommand{\hbx}{\hat{\bm{x}}}
\newcommand{\hby}{\hat{\bm{y}}}

\newcommand{\hp}{\hat{p}}

\newcommand{\hx}{\hat{x}}

\newcommand{\hH}{\hat{H}}

\newcommand{\brho}{\bm{\rho}}

\newcommand{\be}{\bm{e}}

\providecommand{\abs}[1]{\lvert#1\rvert}

\providecommand{\ket}[1]{|#1\rangle}

\providecommand{\brak}[2]{\langle#1|#2\rangle} 
\usepackage[all]{xy}
\begin{document}
\title{Classical entanglement: Oxymoron or resource?}
\author{Andrea Aiello$^{1,2}$}
\email{andrea.aiello@mpl.mpg.de}
\author{Falk T\"{o}ppel$^{1,2,3}$}
\author{Christoph Marquardt$^{1,2}$}
\author{Elisabeth Giacobino$^{1,4}$}
\author{Gerd Leuchs$^{1,2}$}
\affiliation{$^1$ Max Planck Institute for the Science of Light, G$\ddot{u}$nther-Scharowsky-Strasse 1/Bau24, 91058 Erlangen,
Germany}
\affiliation{$^2$Institute for Optics, Information and Photonics, University of Erlangen-Nuernberg, Staudtstrasse 7/B2, 91058 Erlangen, Germany}
\affiliation{$^3$ Erlangen Graduate School in Advanced Optical Technologies (SAOT), Paul-Gordan-Stra{\ss}e 6, 91052 Erlangen, Germany}
\affiliation{$^4$ Laboratoire Kastler Brossel, Universit\'e Pierre et Marie Curie, Ecole Normale Sup\'erieure, CNRS, 4 place Jussieu, 75252 Paris Cedex 05, France}
\date{\today}
\begin{abstract}
In this work we  review and further develop the 
controversial concept of ``classical entanglement'' in optical beams. 
We present a unified theory for different kinds of light beams exhibiting classical entanglement and we indicate several possible extensions of the concept.  Our results shed new light upon the physics at the debated border between the classical and the quantum representations of the world.
\end{abstract}
\maketitle
\section{Introduction}\label{sec1}
A composite physical system, namely one made of at least two identifiable parts, say $A$ and $B$, which are denoted subsystems, can be prepared in such a way that the latter are not independent. In the realm of classical physics this means, for example, that the probability $P(a\in A, b \in B)$ for the events $a,b$ associated to subsystems $A,B$, respectively, cannot be factored as  $P(a\in A, b \in B) = P(a\in A)P(b \in B)$ \cite{Papoulis}. 
Conversely, for a composite quantum system, statistical dependence of the subsystems $A,B$ means that the state vector $\ket{\Psi}$ describing a physical state of the whole system, cannot be decomposed in the tensor product
\begin{align}\label{Int0}
\ket{\Psi} = \ket{\psi_A} \otimes \ket{\psi_B},
\end{align}
where $\ket{\psi_A}$ represents the  state of the subsystem $A$ and  $\ket{\psi_B}$ represents the state of the subsystem $B$.
Here, we are not interested in the deep conceptual implications of Eq. \eqref{Int0} but follow, rather, the ``die-hard pragmatist's'' approach \cite{Espagnat} and denote as \emph{entangled} any state vector 
that does not factorizes as in Eq. \eqref{Int0}; namely,
\begin{align}\label{Int1}
entangled \, = \, non\text{--}separable.
\end{align}

Traditionally, entanglement has been regarded either as a peculiar feature of quantum mechanics or, instead, as a powerful resource especially for quantum information science  \cite{Nielsen}. In this paper we adhere to the latter view  and aim at showing how some potentially useful characteristics of  \emph{quantum} entanglement   can be replicated in \emph{classical} systems. 
%
%
%
%
In fact, our ultimate goal is \emph{not} to replace or simulate entangled quantum systems with classical ones in some actual operations. Instead, the aim is to study how to make quantum  entanglement potentialities accessible to classical physics applications  as recently demonstrated, e.g., in classical polarization metrology \cite{Toppel}.

Thus, the main purpose of this paper is to revisit the concept of the so-called  ``classical entanglement'' in optics \cite{Spreeuw98,Ghose14},  and to present a brief but comprehensive overview of  it. 
We would like to stress that ``classical entanglement'' is \emph{not} substitutive of bona fide quantum entanglement, but is a feature exhibited by some classical systems. 
In a sense, which will become more clear later, the name \emph{classical entanglement} denotes the occurrence of some mathematical and physical aspects of quantum entanglement in classical beams of light. In this sense, classical entanglement should not be confused with 
``entanglement simulations in classical optics'', namely the use of classical fields to reproduce non-classical correlations between distinct measurement apparatuses \cite{Lee02,Fu04}. In any case, classical entanglement does not belong to the rich field of studies denoted by the name ``quantum-classical analogies''  \cite{Manko,Dragoman,Cessa}. A  precise definition of what is usually meant with ``classical entanglement'', will be given in Sect. 2.

As a final important remark, the term ``classical'' in the name \emph{classical entanglement}, indicates the non-quantum nature of the excitation of the electromagnetic field. In this paper, typically, we deal with bright beams of light as, e.g., laser beams. However,  whether the beam is very intense or very weak, is a factor that has not influence upon classical entanglement, as it will be shown in Sect. 2.  
Yet, it should be noticed that single-photon excitations permit only the quantum mechanical representation as Fock states and, therefore, will not be  considered here. However, it has been recently demonstrated that single photons can be prepared in a quantum state entangled with the vacuum \cite{Enk03,Kim03,Bjork04,Kimble08,Kimble09}. Single-photon-vacuum entanglement resembles classical entanglement in that there is only one individual physical system, a single-photon in the quantum case and a single bright beam in the classical one, and two (or more) entangled modes of the electromagnetic field \cite{Cardano13,Karimi14,Pereira14}.  This concept will be further discussed in the next section.

%
%
\section{Two types of quantum entanglement}\label{sec3}
%
%
%
Consider a quantum system $S$ made of two parts, denoted with $S_1$ and $S_2$, which are dubbed ``subsystems''. For example, \emph{two}  particles of mass $m$  constrained to move along a line with coordinates $x_1,x_2$, respectively,  tied to the equilibrium point by two equal springs of  elastic constant $k = m \omega^2$, constitute a composite (bipartite) system whose  dynamics is governed by the  Hamiltonian
$\hH = \hH_1 + \hH_2$, where 
\begin{align}\label{tre10}
\hH_\alpha =  \frac{1}{2 m} \hp_\alpha^2  + \frac{1}{2 }m \omega^2 \hx_\alpha^2, \qquad (\alpha = 1,2).
\end{align}
In this case the two subsystems $S_1,S_2$ are naturally identified with the two particles \cite{Nota1}.

 As a second example, consider now a \emph{single} particle of mass $m$ moving upon the plane   $(x_1,x_2)$  and tied to the equilibrium point $x_1 =0=x_2$ by a spring of  elastic constant  $k = m \omega^2$. This is a two-dimensional harmonic oscillator with Hamiltonian 
\begin{align}\label{tre15}
\hH 
= & \; \frac{1}{2 m} \left( \hp_1^2 + \hp_2^2 \right) + \frac{1}{2 }m \omega^2 \left( \hx_1^2 + \hx_2^2\right) \nonumber \\
 = & \; \hH_1 + \hH_2,
\end{align}
where $\hH_\alpha$ is again given by the expression in Eq. \eqref{tre10}.
In this case the two subsystems $S_1,S_2$ are clearly identified with the two Cartesian coordinates of the single particle. 
Not surprisingly, the Hamiltonian $\hH$ is the same in both cases and the generic state vector $\ket{\Psi}$ satisfying the  Schr\"{o}dinger equation $i  \hbar {\partial  \ket{\Psi}}/{\partial \, t} = \hH \, \ket{\Psi}$, belongs to a Hilbert space $\mathscr{H}$ made as the tensor product of spaces associated to each subsystem:  $\mathscr{H} = \mathscr{H}_1 \otimes  \mathscr{H}_2$.

The fundamental difference between the two cases considered above is that in the first case the two subsystems
are identified with two distinct physical objects, the two particles, which
 can be  spatially separated. Conversely, in the second case there are not two individual physical objects to set apart but only two orthogonal coordinates attached to a single physical object: the sole particle. This simple fact has serious consequences when the state vector $\ket{\Psi}$ is entangled,  namely when $\ket{\Psi} \neq \ket{\psi_1} \otimes \ket{\psi_2}$. In the words of Spreeuw \cite{Spreeuw01}:  
\begin{quote}
``[there is] a profound difference between two
types of entanglement: (i) true, multiparticle entanglement
and (ii) a weaker form of entanglement between different
degrees of freedom of a single particle. Although these
two types look deceptively similar in many respects, only
type (i) can yield nonlocal correlations. Only the type (ii)
entanglement has a classical analogy.''
\end{quote}
In this paper, borrowing from the jargon of the theory of optical coherence functions \cite{Loudon}, 
we denote entanglement of type (i) and (ii)  as \emph{inter}system and \emph{intra}system entanglement, respectively. 
As remarked by Spreuuw, intersystem entanglement can occur only in quantum systems and may lead to the so-called quantum non-locality \cite{Bell,Englert13}, a fundamental aspect of quantum mechanics that should not be confused with quantum entanglement \cite{Brunner05}. Conversely, intrasystem entanglement may appear in both quantum and classical systems and has a local nature by definition because the two or more entangled degrees of freedom are localized within the same physical object. 

 In the last two decades it became clear that  intrasystem entanglement also occurs frequently in  classical optics. In this case  intrasystem entanglement is usually dubbed \emph{classical entanglement} \cite{Luis09,Ghose14,Simon10,Holleczek11,Gabriel11,Nota2}. A typical example thereof is given by a collimated optical beam with \emph{nonuniform} polarization pattern. 
 The electric field of a generic paraxial  beam of light can be   written as $\vE(\brho,z,t) = 2 \operatorname{Re} \left\{\mathbf{U}(\brho,z)\exp[i k (z - c \, t) ]\right\}$, where $\mathbf{U}(\brho,z)$ is the complex amplitude of the field (technically called: analytic signal \cite{MandelBook}),  $\brho = \hbx x + \hby y$ denotes the transverse position vector, $k$ is the wave-number and the axis $z$ is taken along the direction of propagation of the beam.
Then, the analytic signal of a non-uniformly polarized paraxial beam can be represented by a \emph{non-separable} vector function of the form
\begin{align}\label{Int2}
\mathbf{U}(\brho,z) = {\bm{a}}_1  b_1(\brho,z) + {\bm{a}}_2    b_2(\brho,z),
\end{align}
where $\bm{a}_1,\bm{a}_2 $ are two constant vectors perpendicular to the propagation axis $z$,  the functions $b_1(\brho,z),b_2(\brho,z)$ denotes two  distinct solutions of the paraxial equation. 
In this instance the polarization vectors $\bm{a}_1,\bm{a}_2 $ and the spatial mode functions $b_1(\brho,z),b_2(\brho,z)$ describe two independent degrees of freedom, which play the role of the two subsystems in quantum mechanics.  The degrees of freedom are independent in the sense that 
it is possible to assign arbitrary values to the polarization of a paraxial beam of light irrespective of its spatial mode function and vice versa. 

An expression of the form 
 \eqref{Int2} is clearly non-separable, namely it is not possible to rewrite it as the simple product between one constant polarization vector ${\bm{a}} $ and one mode function $b(\brho,z)$: $\mathbf{U}(\brho,z) \neq {\bm{a}}  \, b(\brho,z)$. In this sense, equation \eqref{Int2} has the same mathematical structure  (isomorphism) of a two-qubit  entangled state vector $\ket{\Psi}$ belonging to a bipartite Hilbert space $\mathscr{H} = \mathscr{H}_1 \otimes  \mathscr{H}_2$ of dimension $4$ \cite{Nielsen}.
It is well known that such state $\ket{\Psi}$ can  always be written in terms of a Schmidt decomposition of the form \cite{Ekert,Peres}
\begin{align}\label{tre14}
\ket{\Psi} =  \sqrt{\lambda_1} \, \ket{ u_1}  \ket{ v_1} + \sqrt{\lambda_2} \,\ket{ u_2}  \ket{ v_2} , 
\end{align}
where  $\{\ket{ u_1} ,\ket{ u_2} \}$ and  $\{\ket{ v_1},\ket{ v_2}\}$ are  orthonormal bases for  $\mathscr{H}_1$ and  $\mathscr{H}_2$, respectively, and $\lambda_1 \geq \lambda_2 \geq 0$ are real non-negative coefficients. If the state is normalized to $1$, then $\lambda_1 + \lambda_2 =1$. 
If either $\lambda_1 =0$ or $\lambda_2=0$ the state is factorable and the two subsystems are independent. Vice versa, if $\lambda_1,\lambda_2 \neq 0$, the state vector $\ket{\Psi}$ is  entangled. The amount of entanglement can be quantified by the Schmidt number  (or participation ratio) $K$ defined as:
\begin{align}\label{tre16}
K =  \frac{ \left(\lambda_1 + \lambda_2 \right)^2}{\lambda_1^2 + \lambda_2^2} ,
\end{align}
with $1 \leq K \leq 2$ \cite{Eberly1,Eberly2}. $K=1$ characterizes factorable state vectors, while $K=2$ denotes maximal entanglement occurring whenever $\lambda_1=\lambda_2$.
In a similar manner, it is not difficult to show that a non-separable vector function of the form \eqref{Int2} can always be rewritten as 
\begin{align}\label{tre18}
\mathbf{U}(\brho,z) =   \sqrt{\lambda_1} \,\hat{\bm{u}}_1  v_1(\brho,z) +  \sqrt{\lambda_2} \,\hat{\bm{u}}_2  v_2(\brho,z),
\end{align}
where 
\begin{align}\label{tre17}
\int \mathbf{U}^*(\brho,z)\cdot \mathbf{U}(\brho,z) \, \di^2 \rho = \lambda_1 +\lambda_2,
\end{align}
denotes the total intensity of the beam and the integration extended upon the whole $xy$ plane with  $\di^2 \rho =  \di x \di y$. 
Here $(\hat{\bm{u}}_\alpha,\hat{\bm{u}}_\beta)_P=\hat{\bm{u}}_\alpha^* \cdot \hat{\bm{u}}_\beta = \delta_{\alpha \beta}$, with $\alpha,\beta \in \{1,2\}$, and 
\begin{align}\label{tre19}
(v_\alpha,v_\beta)_S=\int  v_\alpha^*(\brho,z)  v_\beta(\brho,z) \, \di^2 \rho =\delta_{\alpha \beta}.
\end{align}
 In the two equations above $(\hat{\bm{u}}_1,\hat{\bm{u}}_2)_P$ and $(v_1,v_2)_S$ symbolize the scalar product in the polarization (subscript $P$) and in the spatial (subscript $S$) Hilbert spaces $\mathscr{H}_1$ and $\mathscr{H}_2$, respectively. Given the decomposition \eqref{tre18}, one can again formally quantify the amount of ``classical entanglement'' via the Schmidt number $K$ given in Eq. \eqref{tre16} which holds irrespective of the normalization of the state. Therefore, the total intensity of the beam $\lambda_1 + \lambda_2$ does not affect classical entanglement.
\section{Three kinds of classical entanglement}\label{sec3.1}
%
%
Nowadays, three  methods to prepare optical beams exhibiting intrasystem entanglement are quite popular. In all the three cases the goal is to prepare beams of light possessing some properties of entangled  states of two qubits. This is achieved by manipulating two relevant \emph{binary} degrees of freedom of the electromagnetic field, each qubit being encoded in one degree of freedom. 
According to what pair of binary degrees of freedom are chosen, one can have \emph{1.} polarization-position entanglement, \emph{2.} position-position entanglement and \emph{3.} polarization-spatial entanglement. In the remainder of this section we shall illustrate and compare these three kinds of classical entanglement  within the framework of paraxial optics that allows for a unified description of these cases. 
\subsubsection{Polarization-Position entanglement}\label{sec3.1.1}
\begin{figure}[!h]
\begin{center}
\includegraphics[width=6cm]{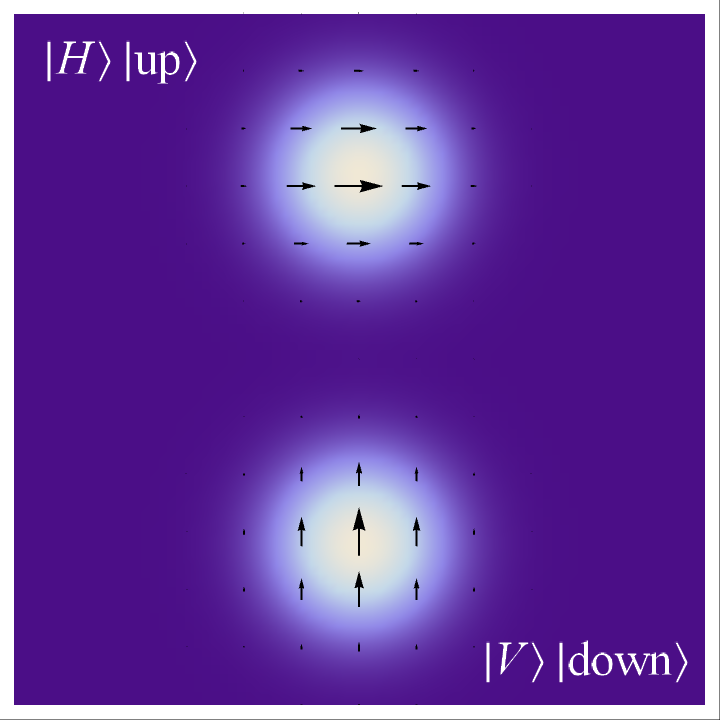}
\caption{\label{fig1}
Polarization and intensity patterns in the transverse plane $z=0$, of the twofold optical beam represented by Eq. \eqref{tre40}. The black arrows denotes the polarization of light.}
\end{center}
\end{figure}

The first example of intrasystem entanglement in optical beams was given by Peres \cite{Peres}, Spreuuw \cite{Spreeuw98} and Cerf \emph{et al.} \cite{Cerf98}. Consider an unpolarized beam of light  passing through a Calcite crystal. 
Crossing the crystal, the beam splits in two beams traveling along two different paths (say ``{u}p'' and ``{d}own''), with orthogonal linear polarization (say, ``\emph{H}orizontal'' and ``\emph{V}ertical''), as shown in Fig. \ref{fig1}.
%
%
%
%
Therefore, the beam can be described in terms of two significant binary degrees of freedom: the polarization and the position of the path. 
As shown in Sect. 2, the paraxial twofold beam exiting the crystal can be  represented by the non-separable vector field
\begin{align}\label{tre20}
\mathbf{U}(\brho,z) = \hbe_H \, U_\text{up}(\brho,z) + \hbe_V  U_\text{down}(\brho,z),
\end{align}
where the polarization vectors are orthogonal by definition $(\hbe_H,\hbe_V)_P = 0$, and $(U_\text{up},U_\text{down})_S = 0$ when the two paths are non overlapping and therefore fully distinguishable. Thus, Eq. \eqref{tre20} has a Schmidt form  analogous to Eq. \eqref{tre18}, with $\lambda_1=\lambda_2=1$ and represents a classical optics analogue of a maximally entangled state of two qubits of the form
\begin{align}\label{tre30}
\ket{H}\ket{\text{up}} + \ket{V}\ket{\text{down}}.
\end{align}

For the sake of definiteness,  let us choose $U_\text{up}(\brho,z) = U(x,y-a\/,z)$ and $U_\text{down}(\brho,z) = U(x,y+a\/,z)$, where $2a\/>0$  quantifies the distances between the two beams and $U(\brho,z)$ denotes any solution of the paraxial equation. By definition, $U(x,y \mp a\/,z)$ represents a beam displaced up and down by $\pm a\/$ along the (vertical) $y$-axis. Thus,  Eq. \eqref{tre20} can be rewritten as
\begin{align}\label{tre40}
\mathbf{U}(\brho,z) = \hbe_H U(x,y-a\/,z) + \hbe_V \, U(x,y+a\/,z).
\end{align}
The orthogonality requirement $(U_\text{up},U_\text{down})_S = 0$ now becomes $I(a\/)=0$, where $I(a\/)$ is the overlap integral
\begin{align}\label{tre50}
I(a\/)=\int U^*(x,y-a\/,z)U(x,y+a\/,z)\, \di x \di y .
\end{align}
This condition is trivially satisfied when the two beams are non overlapping, namely when the functions $U(x,y-a\/,z)$ and $U(x,y+a\/,z)$ have spatially disjoint supports \cite{support} and, therefore, $U_\text{up}^*U_\text{down}=0$, namely:
\begin{align}\label{tre55}
U^*(x,y-a\/,z)U(x,y+a\/,z) = 0.
\end{align}
 For example, for a fundamental Gaussian beam of waist (spot size) $w_0$ and  normalized amplitude 
\begin{align}\label{tre60}
U(\brho,z) = \sqrt{\frac{k L}{\pi}} \frac{1}{z- i L}\exp \left(\frac{i k }{2} \frac{\abs{\brho}^2}{z - i L} \right),
\end{align}
where $L = k w_0^2/2$ is the Rayleigh range of the beam, it is not difficult to show that
\begin{align}\label{tre70}
I(a\/)= \exp \left(- \frac{a\/^2 }{w_0^2/2} \right).
\end{align}
As expected, one obtains $I(a\/) \simeq 0$ only when the separation between the two beams is much bigger than the beam waist: $a\/ \gg w_0$. 
Conversely, when $I(a\/) \neq 0$ the spatial mode functions $U(x,y-a\/,z)$ and $U(x,y+a\/,z)$ are not reciprocally orthogonal and, therefore, Eq. \eqref{tre40} is no longer in a Schmidt form. In this case a new Schmidt decomposition must be performed to bring $U(\brho,z)$ to the form  \eqref{tre18}.
\subsubsection{Position-Position entanglement}\label{sec3.1.2}
\begin{figure}[!h]
\begin{center}
\includegraphics[width=6cm]{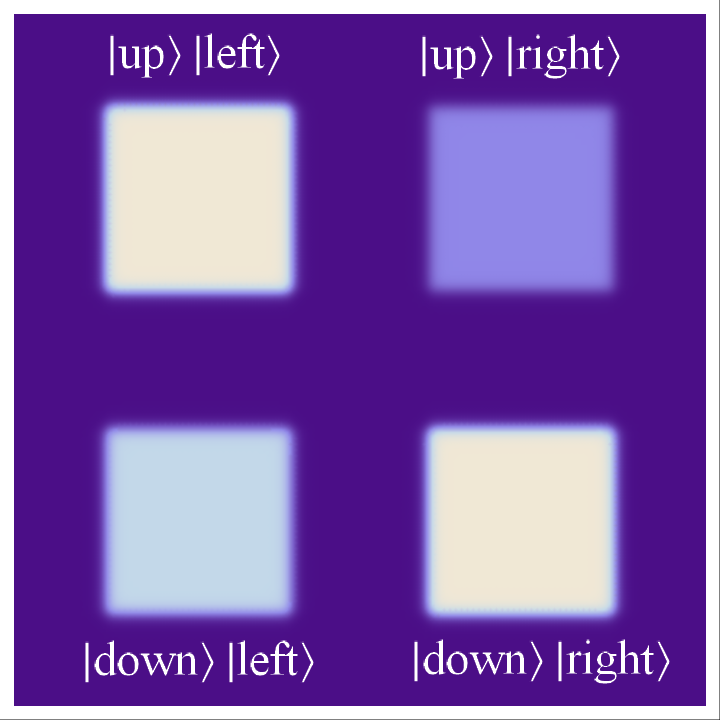}
\caption{\label{fig2}
Illustration of a fourfold optical beam in the transverse plane $z=0$ described by Eq. \eqref{tre90}. The two yellow square spots along the diagonal constitute a twofold beam described by Eq. \eqref{tre130} and represent the entangled vector state \eqref{tre120}.}
\end{center}
\end{figure}

A second way to encode two qubits in optical beams was  proposed by Puentes \emph{et al.} \cite{Puentes04} (a similar method to process optical beams was previously proposed by Caulfield and Shamir  \cite{Shamir} and by Spreeuw and coworkers \cite{Spreeuw02}) and found numerous applications in recent years \cite{Tate05,Francisco06,Francisco08,Goldin10,Goldin11}. The key idea is to encode two qubits in the transverse positions of four non-overlapping beams of light propagating along a common axis, say $z$. In the $xy$-plane of equation $z=0$, these beams form an array of four bright spots  with the same polarization, say $\be$, but different phase and intensity, as shown in Fig. \ref{fig2}.
%
%
%
%
 The first qubit is encoded in the vertical position  (``up'' and ``down'') of such fourfold beam, and the second qubit in the horizontal position (``left'' and ``right'').  Then, the fourfold beam  at $z=0$ can be described by the analytic signal $\mathbf{U}(\brho,0) = \be \, U(x,y)$, where   \cite{Goldin10}
\begin{align}\label{tre90}
U(x,y) = & \; A_{00}U (x+a,y-a)+ A_{01}U (x-a,y-a)\nonumber \\
%
%
& \; + A_{10}U (x+a,y+a)+ A_{11}U (x-a,y+a),
%
%
\end{align}
with $A_{ij} \in \mathbb{C}, \; (i,j = 0,1)$ being numerical constants settling the intensity and the phase of each of the four beams, and the four functions
\begin{multline}\label{tre100}
U (x + (-1)^j a, y -(-1)^i a) \\ = 
 \operatorname{rect} \left(\frac{y -(-1)^i a}{b} \right)\operatorname{rect} \left(\frac{x + (-1)^j a}{b} \right),
\end{multline}
fix the position and the (square) spatial profiles of the beams where, again, $i,j = 0,1$. In Eq. \eqref{tre100},
 $0<b <2 a$ is the horizontal and vertical width of each of the four beams,  and  the rectangle function $\operatorname{rect}(\xi) $ is equal to $1$ for $\abs{\xi} <1/2$,  to $1/2$ for $\abs{\xi} =1/2$ and to $0$ for $\abs{\xi} >1/2$ \cite{Goodman}.

By selecting only two spots along the diagonal $x+y=0$, one achieves the position-position optical beam representation of the two-qubit entangled state 
\begin{align}\label{tre120}
\ket{\text{up}}\ket{\text{left}} + \ket{\text{down}}\ket{\text{right}},
\end{align}
where the spatial separation  $a > b/2$ between the beams guarantees that $\brak{\text{up}}{\text{down}}=0=\brak{\text{left}}{\text{right}}$.
The analytic signal of such twofold beam can be written as
\begin{align}\label{tre130}
U(x,y) = & \; U (x+a,y-a) + U (x-a,y+a) \nonumber \\
= & \; \operatorname{rect}\left(\frac{y - a}{b} \right)  
 \operatorname{rect}\left(\frac{x + a}{b} \right)  \nonumber \\
 & \;+  \operatorname{rect} \left(\frac{y + a}{b} \right) 
 \operatorname{rect} \left(\frac{x - a}{b} \right),
\end{align}
where $\operatorname{rect}\left[({y - a})/{b} \right] \sim \ket{\text{up}}, \; \operatorname{rect}\left[({y + a})/{b} \right] \sim \ket{\text{down}}$ and $ \operatorname{rect}\left[(x + a)/b \right] \sim \ket{\text{left}}, \; \operatorname{rect}\left[(x - a)/b \right] \sim \ket{\text{right}}$.

The functions $U (x+a,y-a)$ and $U (x-a,y+a)$ in Eq. \eqref{tre130} are non-overlapping only if $b<2 a$. In this case the vertical and horizontal positional degrees of freedom are binary  and the Eq. \eqref{tre130} is automatically in a Schmidt form and displays maximum entanglement.
\subsubsection{Polarization-Spatial entanglement}\label{sec3.1.3}
The third method  to achieve intrasystem entanglement in optical beams, exploits polarization and the so-called  first-order spatial modes \cite{Beijersbergen93} of the electromagnetic field, as binary degrees of freedom.
It is a well established result of polarization optics that the polarization vector space  can be represented by the  polarization Poincar\'e sphere \cite{Damask}. It is also known that the vector space formed by the first-order spatial mode can be mapped into a Poincar\'e sphere \cite{Padgett99}.
 The direct product of polarization and spatial vector spaces  contains a subspace spanned by the so-called cylindrically polarized beams of light and can be represented as the direct sum of two ``hybrid'' Poincar\'e spheres \cite{Holleczek10,Milione11}.
In recent years, many fundamental and applied researches upon polarization-spatial entanglement in optical beams have been carried out 
\cite{Oliveira05,Souza07,Luis09,Borges10,Karimi10,Holleczek10,Simon10,Holleczek11,Gabriel11,Qian11,Kagalwala12,Ghose14,Paz14,Toppel,Qian14}.

The analytic signal of the more general paraxial beam in the polarization-spatial  space takes the form
\begin{align}
\label{tre140}
  \mathbf{U}(\brho,z)= & \; A_{00} \be_H U_{10}(\brho,z)+A_{01} \be_H U_{01}(\brho,z) \nonumber \\
& + A_{10}  \be_V U_{10}( \brho,z )+A_{11}   \be_V U_{01}( \brho,z ),
\end{align}
where $A_{ij} \in \mathbb{C}, \; (i,j \in \{0,1\})$ are numerical constants, and $U_{nm}( \brho,z )$ denotes the Hermite-Gauss  solution of the paraxial wave equation of order $N=n+m$ with $N=1$. These solutions are also known as transverse electromagnetic ($\text{TEM}_{nm}$) modes and are orthogonal with respect to the spatial scalar product: $\left(U_{nm}, U_{n'm'} \right)_S= \delta_{n n'} \delta_{m m'}$,
with $n,n',m,m' \in \{0,1,2,\ldots\}$ \cite{SiegmanBook}.

\begin{figure}[!h]
\begin{center}
\includegraphics[width=6cm]{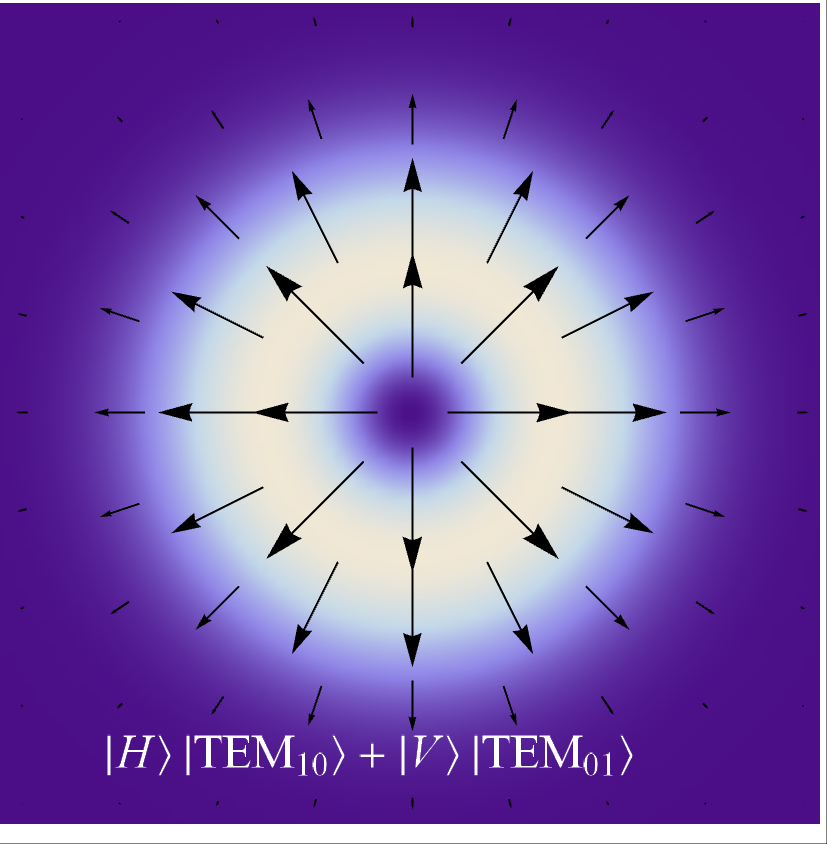}
\caption{\label{fig3}
Polarization (black arrows) and intensity pattern of a radially polarized  optical beam in the transverse plane $z=0$, as described by Eq. \eqref{tre150}. }
\end{center}
\end{figure}

Choosing $ A_{00} = 1 =  A_{11}$ and $ A_{10} = 0 =  A_{01}$ in Eq. \eqref{tre140}, one obtains a representation of the so-called radially polarized beam of light
\begin{align}\label{tre150}
\mathbf{U}( \brho,z )= \be_H  U_{10}( \brho,z ) + \be_V  U_{01}( \brho,z ),
\end{align}
illustrated in Fig. \ref{fig3}.
%
%
%
%
\\
The beam in Eq. \eqref{tre150} is automatically in a Schmidt form and furnishes the polarization-spatial optical  representation of the two-qubit entangled state 
\begin{align}\label{tre160}
\ket{H}\ket{\text{TEM}_{10}} + \ket{V}\ket{\text{TEM}_{01}}.
\end{align}

Non-uniformly polarized beams of light exhibition classical entanglement have recently found practical applications in quantum information \cite{Gabriel11} and polarization metrology \cite{Simon10,Toppel}.
\subsubsection{Comparison}\label{sec3.1.4}
The three kinds of classically entangled optical beams considered in this section have a quite different nature.
First of all,  both polarization-position and polarization-spatial entanglement are consequences of a natural partition between \emph{different} degrees of freedom, namely \emph{polarization} and \emph{position/spatial}. Conversely, position-position entanglement occurs because of an arbitrarily chosen partition of the $\mathbb{R}^2$ plane, being the two binary positional degrees of freedom of the same type. This means, for example, that it is possible to represent the vector field \eqref{tre130} in a separable form by simply choosing a $45^\circ$-rotated Cartesian reference frame.

Now, let us compare  polarization-position (PP) and polarization-spatial (PS) entanglement. 
To begin with, it is clear that Eq. \eqref{tre150} has the same form of Eq. \eqref{tre40}. 
Both expressions are written as a Schmidt sum. Each term in the sums is given by the product of a polarization vector times a scalar function. The two scalar functions in Eq. \eqref{tre40} are orthogonal to each other and the same applies to the two scalar functions in Eq. \eqref{tre150}. However, and here is the profound difference, the functions in Eq. \eqref{tre40} are ``trivially'' orthogonal simply because they are non-zero in different spatially disjoint regions,  as shown by Eq. \eqref{tre55}, i.e., the beams are non overlapping. Conversely, the functions $U_{10}$ and $U_{01}$ in Eq. \eqref{tre150} have the same support \cite{support} and $\left(U_{10},U_{01} \right)_S=0$, although 
\begin{align}\label{tre180}
U_{10}^*( \brho,z )  U_{01}( \brho,z ) \neq 0.
\end{align}
Therefore, in the polarization-spatial entanglement there is a \emph{single} beam of light, the radially polarized one, encoding both qubits. This is very different from the polarization-position case where one needs \emph{two}  spatially separated (therefore, fully distinguishable) beams to encode two entangled qubits.

This concept may be further clarified  noticing that Eq. \eqref{tre150} represents a \emph{coherent} superposition of beams with orthogonal polarization, while Eq. \eqref{tre40} represents, de facto, an \emph{incoherent} superposition of orthogonally polarized beams.
To be more quantitative, we may calculate the covariance matrix \cite{MandelBook} of both PP and PS beams defined as
\begin{align}\label{tre190}
J^\text{PX} = \int \mathcal{J}^\text{PX}( \brho,z ) \, \di^2 \rho,
\end{align}
with $\text{X} \in \{ \text{P,S}\}$, and  
\begin{align}\label{tre200}
\mathcal{J}^\text{PP}( \brho,z ) = & \; \left[
\begin{array}{cc}
\abs{U_\text{up}}^2  & U_\text{up}U_\text{down}^* \\
U_\text{up}^*U_\text{down}  &  \abs{U_\text{down}}^2
\end{array} \right] \nonumber \\
\Leftrightarrow & \; \left[
\begin{array}{cc}
\abs{U_\text{up}}^2  & 0 \\
0  &  \abs{U_\text{down}}^2
\end{array} \right] ,
\end{align}
where the last line is a straightforward consequence of Eq. \eqref{tre55}, and
\begin{align}\label{tre205}
\mathcal{J}^\text{PS}( \brho,z ) = & \; \left[
\begin{array}{cc}
\abs{U_{10}}^2  & U_{10}U_{01}^* \\
U_{10}^*U_{01}  &  \abs{U_{01}}^2
\end{array} \right].
\end{align}
Spatial integration in Eq. \eqref{tre190} has the physical meaning of disregarding the position/spatial degrees of freedom. It is the analogous of the ``trace'' operation in quantum mechanics, with respect to the unobserved subsystem.
 A straightforward calculation shows that  in both cases it has $J^\text{PX} = I_2$, where $I_2$ denotes the $2 \times 2$ identity matrix. This is expected because both beams represents maximally entangled states \cite{Abouraddy01, Abouraddy02} and the corresponding covariance matrix must describe completely unpolarized light. However, the diagonal form of Eq. \eqref{tre200} reveals that Eq. \eqref{tre40} is in some sense more similar to an incoherent superposition already before integration.
In this respect, polarization-spatial entanglement is the ``closest'' one, amongst the three types of entanglement considered here, to quantum entanglement.
%
%
%
\section{Outlook: from $2$ qubits to $3$ qubits entanglement and more}\label{sec5}
In the case of polarization-spatial entanglement, we have considered each Hermite-Gauss mode $U_{nm}( \brho,z )$ as a single function. However, from the case of position-position entanglement we have learned that the Cartesian coordinates $x$ and $y$ may be also considered as independent degrees of freedom.  In this section, we combine these two concepts to build optical beam representations of tripartite states of qubits, each party being associate to a specific degree of freedom.

We begin with the simple observation that an Hermite-Gauss mode can be always factorized as 
\begin{align}\label{20}
U_{nm}( \brho,z ) = u_{n}(x,z) u_{m}(y,z),
\end{align}
where $n,m = 0,1,2,\ldots$ . This means, for example, that the radially polarized beam \eqref{tre150}, can be rewritten in the form
\begin{align}\label{30}
\!\!\!\!\mathbf{U}( \brho,z )= \be_H  u_{1}(x,z)u_{0}(y,z) + \be_V \, u_{0}(x,z)u_{1}(y,z),
\end{align}
which is  isomorphic to the three-qubit state vector
\begin{align}\label{40}
 \ket{0}_\text{p} \ket{1}_\text{x} \ket{0}_\text{y} +  \ket{1}_\text{p} \ket{0}_\text{x} \ket{1}_\text{y} ,
\end{align}
where $\be_H \sim \ket{0}_\text{p}, \; \be_V \sim \ket{1}_\text{p}$, $u_{1}(x,z) \sim \ket{1}_\text{x}, \; u_{0}(y,z) \sim \ket{0}_\text{y}$ and $u_{0}(x,z) \sim \ket{0}_\text{x}, \; u_{1}(y,z) \sim \ket{1}_\text{y}$, and the label ``p'' stands for ``polarization''.
The state vector \eqref{40} lives in a Hilbert space with $2 \times 2 \times 2$ dimensions, namely it represents a tripartite system. Of course, a state of the form \eqref{40} can be easily generalized to a vector state acting in a $2 \times N \times M$ Hilbert space. The idea of considering entanglement between $x$ and $y$ Cartesian coordinates in paraxial beams has been also recently exploited by Agarwal and coworkers \cite{Chowdhury2013}.
A tripartite representation of an optical beam like the one in Eq. \eqref{30} can be used, for example, to implement a quantum-like teleportation scheme. Other examples of possible applications are given below.
\subsection{GHZ state}
\begin{figure}[!h]
\begin{center}
\includegraphics[width=7cm]{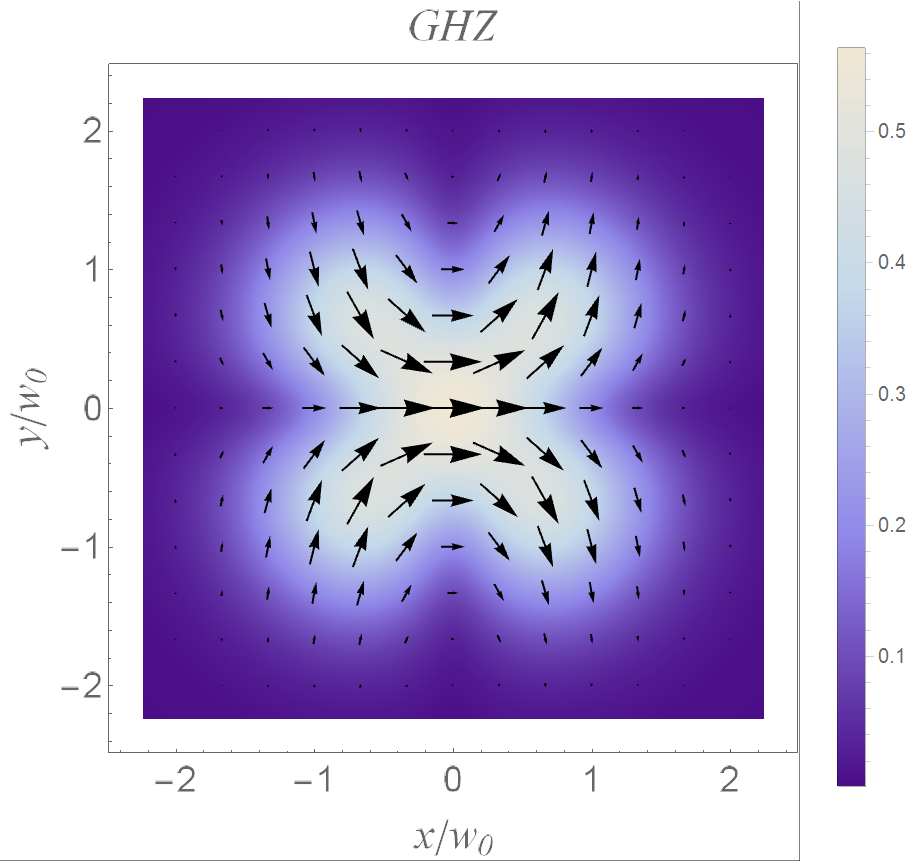}
\caption{\label{fig4}
Polarization (black arrows) and intensity pattern of the non-uniformly polarized  beam  described by Eq. \eqref{60} analogous to the quantum GHZ vector state \eqref{50}. }
\end{center}
\end{figure}

The state vector given in Eq.  \eqref{40} as a form similar to the so-called GHZ state \cite{GHZ}:
\begin{align}\label{50}
|\mathrm{GHZ}\rangle = \frac{1}{\sqrt{2}} \big( |000\rangle + |111\rangle \big).
\end{align}
In the language of classical optics, the lowest-order implementation of the state \eqref{50} is given by the beam 
\begin{align}\label{60}
\mathbf{U}_\mathrm{GHZ}(\brho,z) = \frac{1}{\sqrt{2}} \Big[ \be_H U_{00}(\brho,z) +\be_V  U_{11}(\brho,z) \Big].
\end{align}
 The beam represented by Eq. \eqref{60} clearly has a non-uniform polarization pattern, as shown in Fig. \ref{fig4}.
%
%
%
%
%
\subsection{W state}
Another famous tripartite quantum state is the so-called W state \cite{W}:
\begin{align}\label{70}
|\mathrm{W}\rangle = \frac{1}{\sqrt{3}} \big( |001\rangle +|010\rangle +|100\rangle \big),
\end{align}
which has the same mathematical structure of the following classical beam:
\begin{align}\label{80}
 & \! \! \!\! \mathbf{U}_\mathrm{W}(\brho,z) \nonumber \\
&\! \! \! = \frac{1}{\sqrt{3}} \Big[ \be_H U_{01}(\brho,z) + \be_H U_{10}(\brho,z) +\be_V \,U_{00}(\brho,z) \Big]. 
\end{align}
Also $\mathbf{U}_\mathrm{W}(\brho,z) )$ describes a beam with a non-uniform polarization pattern, as illustrated in Fig. \ref{fig5} below:
%
%
\begin{figure}[!h]
\begin{center}
\includegraphics[width=7cm]{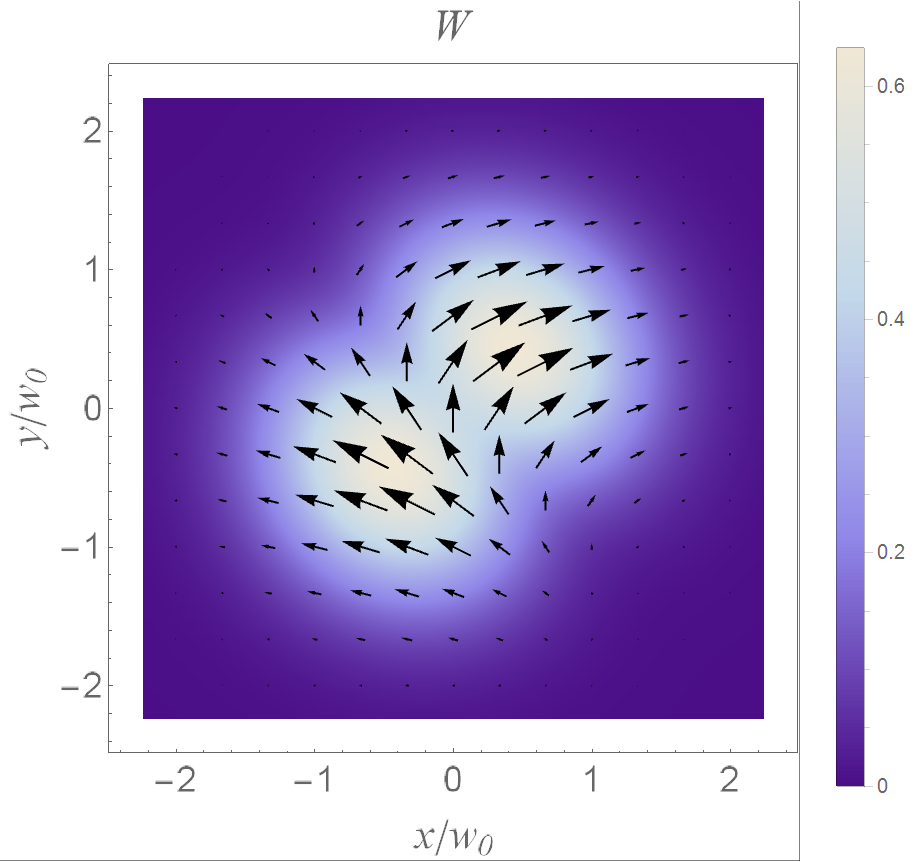}
\caption{\label{fig5}
Polarization (black arrows) and intensity pattern of the non-uniformly polarized  beam  described by Eq. \eqref{80} corresponding to the quantum W vector state \eqref{70}. Differently from the pattern shown in Fig. \ref{fig4}, this beam does  not exhibit Cartesian symmetry.}
\end{center}
\end{figure}
%
%
%
\subsection{NOON states}
\begin{figure}[!h]
\begin{center}
\includegraphics[width=10cm]{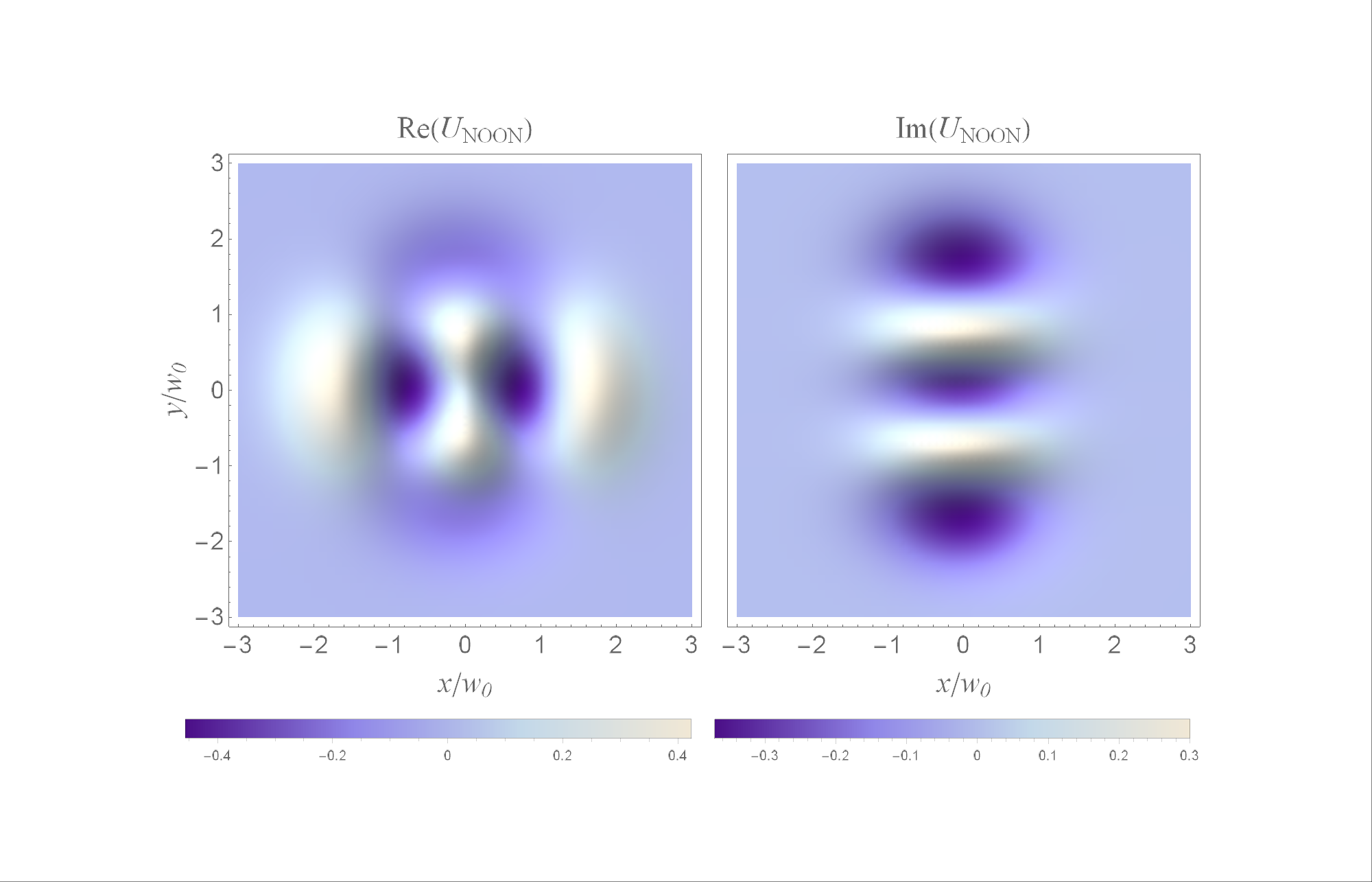}
\caption{\label{fig6}
Real and imaginary parts of the ``NOON'' beam field  described by Eq. \eqref{100} evaluated for $N=4$ and $\theta = \pi/3$.}
\end{center}
\end{figure}

Another class of entangled quantum states that can be encoded in paraxial beams of light, are the so-called generalized NOON states \cite{NOON}:
\begin{align}\label{90}
    |\text{NOON}\rangle = \frac{1}{\sqrt{2}} \big( |N, 0\rangle + e^{iN \theta }|0, N\rangle \big).
\end{align}
The classical optics representation of \eqref{90} is a purely scalar superposition of HG beams: 
\begin{align}\label{100}
U_\mathrm{NOON}(\brho,z) = \frac{1}{\sqrt{2}} \big[U_{N0}(\brho,z) + e^{iN \theta } U_{0N}(\brho,z) \big].
\end{align}
The real and imaginary parts of the analytic signal given in Eq. \eqref{100} are shown in Fig. \ref{fig6} below for $N=4$ and $\theta = \pi/3$: \\ 
%
%
%
%
\\
Super-resolution and super-sensitivity (see, e.g., \cite{White} for a proper definition of the two terms), are intriguing properties of the NOON states that  could be investigated using classical beams of light of the form \eqref{100}. While it is known that super-resolution can be achieved with classical light \cite{White,Sciarrino}, the question whether super-sensitivity could be obtained by classical beams of the form \eqref{100}, is perfectly open. Moreover, these optical beams could furnish a good ``laboratory'' to study decoherence in Schr\"{o}dinger cat states under easily controllable conditions.
\section{Summary}\label{sec6}
The seemingly oxymoronic name ``classical entanglement'' actually denotes the occurrence of some typical quantum mechanical features in classical systems and it should not be regarded as a substitute for quantum entanglement.
In this article we have studied  classical entanglement exhibited by optical beams prepared in three different manners. Within the context of paraxial optics, we have been able to provide a  theory unifying the representation of these three kinds of  beams. We also have classified  the latter according to what pair of binary degrees of freedom of the light is chosen to encode the ``entangled qubits''. Moreover, we have demonstrated that despite of the formal similarity between the mathematical expressions for the 
 beams in all the three cases, the physical characteristics of the light (coherent or incoherent) may be very different. Finally,  we have suggested a few ideas  about how to enlarge the already rich phenomenology of classical entanglement.
%
%
%
%

%
%

\begin{thebibliography}{40}
\bibitem{Nota1}
In this qualitative discussion we are intentionally oversimplifying the situation in order to keep simple the language. More properly, one should identify  the two subsystems with the two harmonic oscillators.
\bibitem{Nota2}
The names ``nonquantum entanglement'' \cite{Simon10}  and ``structural inseparability'' \cite{Holleczek11,Gabriel11} are also sometimes used.
\bibitem{support} Let us briefly remind the concept of support of a function. Let $f: X \to \mathbb{C}$ be  a complex-valued function defined in the arbitrary domain  $X$. The  support of $f$, written $\operatorname{supp}(f)$, is the set of points in $X$ where $f$ is non-zero: $\operatorname{supp}(f) = \{x \in X \,|\, f(x)\ne 0\}$.
\end{thebibliography}
\end{document}